\documentclass{PoS}

\title{Inhomogeneous phases near the chiral critical point in NJL-type models}

\ShortTitle{Inhomogeneous phases near the chiral critical point in NJL-type models}

\author{\speaker{Dominik Nickel}\\%
        Massachusetts Institute of Technology\\
        E-mail: \email{dnickel@mit.edu}}

\abstract{
The role of inhomogeneous phases in the phase diagram of the Nambu--Jona-Lasinio (NJL) and the quark meson (QM) model is examined.
By means of a generalized Ginzburg-Landau (GL) expansion it is concluded that the critical point in the mean-field phase diagram of the NJL model is in fact a Lifshitz point where homogeneous spontaneously broken, inhomogeneous and restored phases meet.
This picture is confirmed by a mean-field calculation for inhomogeneous phases with a one-dimensional modulation.
For the latter it is shown that recent results within lower dimensional models can be extended to 3+1 dimensions.
Also the respective phase diagram for the QM model is presented.

\vspace{-16.3cm}\parbox{\textwidth}{\flushright\large\rm \hfill MIT-CTP 4061}\vspace{16.3cm}
}

\FullConference{5th International Workshop on Critical Point and Onset of Deconfinement\\
		 June 8-12, 2009\\
		 Brookhaven National Laboratory, Long Island, New York, USA}

\begin{document}

\section{Introduction and summary}
\noindent
Until today the phase diagram of quantum chromodynamics (QCD) is subject to intense theoretical and experimental investigations (for dedicated reviews see Ref.~\cite{reviews}).
Since ab initio calculations are limited to small net-baryon densities, possible scenarios at moderate temperatures and densities are often discussed within NJL-type models\footnote{We refer to NJL-type models as models that at least in the applied approximation reduce to the NJL model, possibly extended by additional point-like interactions,  on a technical level. This includes e.g. simplified ans\"atze for the gluon interaction, the use of different regularizations, the instanton liquid model and the quark-meson model.},  typically within mean-field approximation.

\noindent
Here we limit ourself to results for inhomogeneous ground states in the NJL model~\cite{Nambu:1961tp} and QM model~\cite{GellMann:1960np}.
These are characterized by a spatially varying order parameter and have been discussed for QCD at least in the large $N$ limit, where they are expected to form the ground state at sufficiently high densities~\cite{Deryagin:1992rw,Shuster:1999tn,Park:1999bz}. Related to this they show up in holographic models~\cite{Rozali:2007rx} and in the quarkyonic matter picture~\cite{McLerran:2007qj} that suggests a similar structure for QCD.
The investigations of these phases is however limited, mainly because they are technically much more involved.
Within the NJL model such phases have been analyzed at vanishing temperatures applying further truncations~\cite{Rapp:2000zd} as well as for the so-called chiral density wave~\cite{Sadzikowski:2000ap,Nakano:2004cd}. In the latter case the order parameter is assumed to be a plane wave and it can be solved on mean-field level for vanishing current quark masses.
Recently, also the effect of small current quark masses has been discussed for this ansatz~\cite{Maedan:2009yi}.

\noindent
Following Ref.~\cite{Nickel:2009ke} we first address inhomogeneous phases by means of a generalized GL expansion, i.e. by an expansion of the thermodynamic potential as an effective action in the order parameters as well as in gradients acting on the order parameter.
This is a systematic expansion on top of a mean-field approximation in the vicinity of a second order phase transition and therefore especially suited to explore the region around the critical point (CP).
For the NJL model it then turns out that the CP is in fact a Lifshitz point, where the first order phase transition in the phase diagram of the NJL model is replaced by two second order phase transition lines that border an inhomogeneous phase and these two transition lines meet at the CP.

\noindent
In order to give a better picture of this finding and in particular to estimate the importance of inhomogeneous phases in the phase diagram also away from the CP, we then focus on a complete mean-field calculation, following Ref.~\cite{Nickel:2009wj}, and consider inhomogeneous ground states that form lattices of domain-wall solitons.
The key observation here is that for the case of one-dimensional modulations in the order parameter, the problem can be reduced to an analogues problem in the 1+1 dimensional (chiral) Gross-Neveu (GN) model.
For this model inhomogeneous phases have been investigated for the large $N$ limit~\cite{Schnetz:2004vr,Schnetz:2005ih,Thies:2006ti,Basar:2008im,Basar:2009fg}, which technically corresponds to a mean-field approximation, and basically all ground-states have been classified at least for the chiral limit~\cite{Correa:2009xa}. For the case of the GN model it is furthermore possible to introduce finite quark masses and to study their effect on the structure of the phase diagram~\cite{Schnetz:2005ih}.
The possibility to use solutions from lower dimensional models is mainly due to the structure of the mean-field Hamiltonian, which is of Dirac-type, and a similar procedure is e.g. not possible in (color-)superconductors. As a consequence the investigation of the latter is much more tedious~\cite{Nickel:2008ng}.

\noindent
Being able to investigate the role of inhomogeneous phases in the phase diagram of the NJL model at least for one-dimensional modulations, we can confirm the picture obtained by the GL expansion for the vicinity of the CP and the absence of a first order phase transition line in the phase diagram\footnote{This picture could of course be modified by the inclusion of color-superconducting phases.}. Furthermore we analyze the relation between inhomogeneous phases and e.g. the strength of the first order phase transition (present in the case of homogeneous phases), explore the role of finite current quark masses and discuss results in the QM model. The purpose for the latter is twofold: On the one hand we would like to extend the analysis to a larger class of models in general, on the other hand the regularization of the NJL model for inhomogeneous phases is non-trivial with regard to a combined vacuum and QCD phase diagram phenomenology.

\section{Generalized Ginzburg-Landau expansion}
\noindent
Following Ref.~\cite{Nickel:2009ke},  we first concentrate on the two-flavor NJL model given by the Lagrangian
\begin{eqnarray}
\mathcal{L}
&=&
\bar{\psi}
\left(
i\gamma^\mu \partial_\mu
-
\hat{m}
\right)
\psi
+
G_s
\left(
\left(\bar{\psi}\psi\right)^2
+
\left(\bar{\psi}i\gamma^5\tau^a\psi\right)^2
\right)
\,,
\end{eqnarray}
where $\psi$ is the $4N_f N_c$-dimensional quark spinor for $N_f=2$ flavors and $N_c=3$ colors, $\gamma^\mu$ are Dirac matrices, $G_s$ is the scalar coupling and $\hat{m}$ the mass matrix for degenerate quarks with current quark mass $m$. For $N_f=2$ the matrices $\tau^a$ are the conventional Pauli matrices.
In mean-field approximation with $\langle\bar{\psi}\psi\rangle=-\frac{1}{2G}M({\bf x})$ and  $\langle\bar{\psi}i\gamma^5\tau^a\psi\rangle=0$, the  Lagrangian gets replaced by the bilinear functional
\begin{eqnarray}
\mathcal{L}_{MF}
&=&
\bar{\psi}(i\gamma^\mu \partial_\mu - M({\bf x}))\psi -\frac{(M({\bf x})-m)^2}{4G}
\,.
\end{eqnarray}
In the case of a periodic condensate with Wigner-Seitz cell $V$ and using the imaginary-time formalism, we therefore obtain for the mean-field thermodynamic potential as an effective action in the order parameter
\begin{eqnarray}
\label{eq:Omega1}
\Omega(T,\mu;M({\bf x}))
&=&
-\frac{T}{V}\ln 
\int \mathcal{D}\bar{\psi}\mathcal{D}\psi \exp\left(\int_{x\in [0,\frac{1}{T}]\times V} (\mathcal{L}_{MF}+\mu \bar{\psi}\gamma^0 \psi)\right)
\nonumber\\
&=&
-\frac{T}{V}
\mathrm{Tr}_{D,c,f,V} \, \mathrm{Log}\left(S^{-1}\right)
+
\frac{1}{V}\int_V
\frac{(M({\bf x})-m)^2}{4G}
+\mathrm{const.}
\,,
\end{eqnarray}
with inverse propagator
\begin{eqnarray}
S^{-1}(x,y)
&=&
(i\gamma^\mu \partial_\mu - M({\bf x}))\delta^{(4)}(x-y)
\,.
\end{eqnarray}
Since the evaluation of the thermodynamic potential for an arbitrary function is non-trivial due to the functional logarithm, we first aim at an expansion in the order parameter $M({\bf x})$.
Setting $m=0$ for simplicity and
substracting the leading order corresponding to the thermodynamic potential of the unbroken phase, we formally arrive at
\begin{eqnarray}
\Delta\Omega(T,\mu;M({\bf x}))
&=&
-\frac{T}{V}
\sum_{n>0}\frac{1}{n}\mathrm{Tr}_{D,c,f,V}\left(S_0 M\right)^n
+
\frac{1}{V}\int_V
\frac{M({\bf x})^2}{4G}
\,.
\end{eqnarray}
Here we have introduced the bare propagator $S_0=S\vert_{M({\bf x})=0}$ and  a short hand notation for
\begin{eqnarray}
\mathrm{Tr}_{D,c,f,V}\left(S_0 M\right)^n
&=&
\int_{x}
\int_{x_2}
\dots
\int_{x_n}
\mathrm{Tr}_{D,c,f}
\left(
M({\bf x})
S_0(x,x_2)
M({\bf x}_2)
\dots
M({\bf x}_n)
S_0(x_n,x)
\right)
\,.
\nonumber\\
\end{eqnarray}
The domain of integration for $x$ is $[0,\frac{1}{T}]\times V$ and $[0,\frac{1}{T}]\times\mathbb{R}^3$ for $x_2,\dots,x_n$.
In the chiral limit the expressions for odd values of $n$ vanish. Furthermore, in order to arrive at a local functional, we can expand the condensate around ${\bf x}$ as
\begin{eqnarray}
M({\bf x}_n)
&=&
\sum_{\vert\alpha\vert>0}\frac{1}{\alpha!}D^{\alpha}M({\bf x})({\bf x}_n-{\bf x})^{\alpha}
\end{eqnarray}
and can extract the GL functional to any desired order in gradients and order parameter.
Neglecting possible issues with the regularization for the moment, we can go to momentum space using
$
S_0(x,y)
=
T\sum_n \frac{d^3 p}{(2\pi)^{3}} ({p_\mu \gamma^\mu})^{-1}\exp(ip(x-y))
\,,
$
where $p_0 = (2n+1)\pi T$.
Treating the magnitude of the order parameter and the gradients to be of the same order, it is then a tedious but straightforward exercise to work out 
\begin{eqnarray}
\label{eq:OmegaGL}
\Omega_{GL}(T,\mu;M({\bf x}))
&=&
\frac{\alpha_2}{2}
M({\bf x})^2
+
\frac{\alpha_4}{4}
\left(M({\bf x})^4 + (\nabla M({\bf x}))^2\right)
\nonumber\\&&
+
\frac{\alpha_6}{6}
\left(
M({\bf x})^6
+
5(\nabla M({\bf x}))^2 M({\bf x})^2
+
\frac{1}{2}(\Delta M({\bf x}))^2
\right)
\,,
\nonumber\\
\end{eqnarray}
where
\begin{eqnarray}
\alpha_n
&=&
(-1)^{\frac{n}{2}} 4 N_f N_c 
T\sum_n
\int_{reg.} \!\! \frac{d^3 p}{(2\pi)^{3}}\,\,
\frac{1}{((\omega_n+ i\mu)^2+p^2)^{\frac{n}{2}}}
+
\frac{\delta_{2n}}{2G}
\,.
\end{eqnarray}
This expression has to be taken with some caution as the model is non-renormalizable. As a result usually a regularization, as part of the phenomenological model, is introduced. Due to this ad hoc procedure a generalization of the regularization to inhomogeneous phases is often not unique. We could therefore take the pragmatic viewpoint that a generalization of any such ad hoc regularization procedure to inhomogeneous phases is assumed to be such that total derivative terms (which arise in the calculation of the GL functional) vanish.
An alternative approach is a regularization scheme that does not rely on an homogeneous ground state, e.g. a propertime regularization for the functional logarithm in Eq.(\ref{eq:Omega1}).
In this case it is possible to show that no additional complication due to ultraviolet divergencies arise.

\noindent
The most interesting feature of the GL functional is that the coefficient of the $M({\bf{x}})^4$-term is equal to the $(\nabla M({\bf{x}}))^2$-term. As discussed below this will allow for inhomogeneous phases in the regime where $\alpha_4<0$.
It is also worth noting that the GL functional takes a similar form as in the one-dimensional GN model~\cite{Schnetz:2004vr,Thies:2006ti}.
This is not the case in superconductors~\cite{Buzdin:1997a} where the underlying dynamics is different, namely coming from particle-particle and hole-hole scattering near the Fermi surface instead of particle-hole scattering in presented case.

\noindent
With the generalized GL functional at hand we can explore the vicinity near the CP, which is defined by $\alpha_2=\alpha_4=0$ and
$\alpha_6>0$. Without giving a specific choice of model parameters, we assume that such a point in the phase diagram exists and focus on the effect of the gradient terms. Specific examples will be subject of the following section.
For $\alpha_4>0$ we have a  conventional second order phase transition at $\alpha_2=0$ between a dynamically broken ($\alpha_2<0$) and a restored phase ($\alpha_2>0$). Limiting to homogeneous phases this phase transition is rendered first order when going through the CP into the regime where $\alpha_4<0$. The transition line for this case is given by $\alpha_4=-\sqrt{16\alpha_2\alpha_6/3}$.
In this case, however, also inhomogeneous phases can be expected since the $(\nabla M({\bf{x}}))^2$-term is negative and a curvature in the order parameter can therefore lead to a gain in free-energy.

\begin{figure}[h]
\begin{center}
\includegraphics[width=7cm]{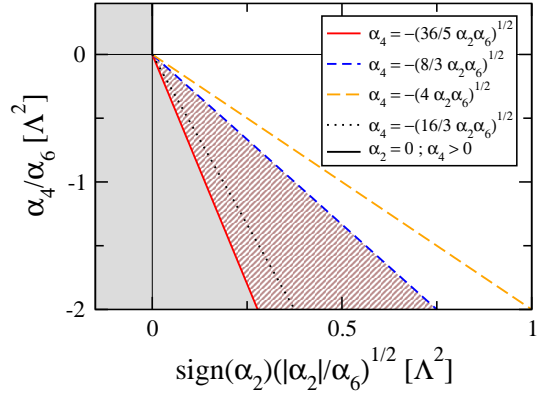}
\caption{
Pictorial presentation of the phase diagram in terms of the GL coefficients: The gray domain corresponds to the homogeneous dynamically broken ground state, the shaded gray to the solitonic ground state (at least when restricting to one-dimensional modulations in the order parameter), whereas in the transparent domain the unbroken phase is preferred. $\Lambda$ is an arbitrary scale. Also stated are various lines discussed in the text.
}
\label{fig:fig1}
\end{center}
\end{figure}

\noindent
To explore this possibility further we limit ourself to a one-dimensional modulation, i.e. $M({\bf{x}})=M(z)$. The solutions to $\frac{\delta}{\delta M}\Delta\Omega = 0$ are actually known from the investigation of one-dimensional models~\cite{Basar:2008im,Buzdin:1997a}. They are expressed (up to an arbitrary shift) in terms of the elliptic Jacobi $\mathrm{sn}$-function as
\begin{eqnarray}
M_{1D}(z)
&=&
\sqrt{\nu} q \, \mathrm{sn}(q z,\nu)
\,,
\end{eqnarray}
where $\nu\in[0,1]$ and $q$ being a scale related to the maximum of $M_{1D}(z)$ and the extension of a soliton in the chosen $z$ direction (both scales are related in our case).
For $\nu=1$ we have $M_{1D}(z) = q \tanh(q x)$, i.e. a single soliton and for $\nu\rightarrow 0$ the shape becomes more and more sinusoidal albeit the amplitude also goes to zero.
From previous investigations it is known that when increasing $\alpha_2$ from zero we reach a second order phase transition into an inhomogeneous phase with $q=M_0$ and $\nu=1$. At this point the free-energy of a single soliton becomes negative leading to its formation.
By using $M_0$ known from the homogeneous case and checking where $\frac{d}{d\nu}\Delta\Omega\vert_{M({\bf{x}})=M_{1D}(z)}$ changes sign at $\nu = 0$, we obtain $\alpha_4 = -\sqrt{\frac{36}{5}\alpha_2 \alpha_6}$ for this point.
We arrive at the onset of infinitely far separated solitons. Further increasing $\alpha_2$ decreases $\nu$ until it reaches zero. Since $q$ stays finite the overall magnitude of $M_{1D}(z)$ given by $\sqrt{\nu}q$ then vanishes and we find a second order phase transition to the unbroken phase.

\noindent
In case of a second order phase transition from the inhomogeneous phase to the unbroken phase, the value of $\alpha_4$ in terms of $\alpha_2 \alpha_6$ is actually universal also for higher dimensional modulations of the order parameter.
Since in this case $M({\bf{x}})$ is parametrically small, we can neglect non-quadratic terms in the GL functional.
Consequently the variation $\frac{d}{d M}\Delta\Omega$  leads to a linear partial differential equation.
We can then optimize the value of $\alpha_4$ by varying the momentum ${\bf{q}}$ of the Fourier components of $M({\bf{x}})$ and find $\alpha_4 =-\sqrt{ \frac{8}{3}\alpha_2 \alpha_6}$ for the transition line where $\vert{\bf{q}}\vert = \sqrt{-\frac{3\alpha_4}{2\alpha_6}}$.

\noindent
The modification of the phase diagram in the vicinity of the CP is also illustrated in Fig.~\ref{fig:fig1}. Neglecting the possibility of inhomogeneous phases, the black dotted line shows the first order phase transition when limiting to homogeneous phases.
The shaded domain then depicts where inhomogeneous phases are energetically preferred. It is enclosed by two second-order phase transition lines that meet at the CP, and it covers the first order phase transition line.

\noindent
We do not want to address the general question whether an inhomogeneous phase with a higher dimensional modulation could become favored in the vicinity of the CP, but it may very well be that the one-dimensional modulations are generally preferred close to the CP as numerically confirmed in Ref.~\cite{Houzet:1999a} for the analogous case of inhomogeneous phases in paramagnetic superconductors.

\section{Phase diagrams allowing for inhomogeneous ground states with a one-dimensional modulation}
\noindent
In order to determine the mean-field thermodynamic potential without further approximations, we evaluate Eq.(\ref{eq:Omega1}) to
\begin{eqnarray}
\Omega(T,\mu;M({\bf x}))
&=&
-\frac{2T N_c}{V}
\sum_{n}
\mathrm{Tr}_{D,V} \, \mathrm{Log}\left(\frac{1}{T}\left(i\omega_{n}+\tilde{H}_{MF}-\mu\right)\right)
+
\frac{1}{V}\int_V
\frac{(M({\bf{x}})-m)^2}{4G_s}
\,,
\nonumber\\
&=&
-\frac{2T N_c}{V}
\sum_{E_{n}}
\ln\left(2\cosh\left(\frac{E_{n}-\mu}{2T}\right)\right)
+
\frac{1}{V}\int_V
\frac{( M({\bf{x}})-m)^2}{4G_s}
+
\mathrm{const.}
\,,
\end{eqnarray}
where we introduced the energy-spectrum $\{E_n\}$ of the Dirac-type Hamiltonian
\begin{eqnarray}
\label{eq:hamiltonian0}
{H}_{MF}
&=&
-i\gamma^0\gamma^{i}\partial_{i} + \gamma^0 M({\bf{x}})
\,.
\end{eqnarray}

\noindent
Limiting to order parameters with a one-dimensional modulation $M({\bf{x}})=M(z)$, we can make use of two connected properties. First, we use the conserved momentum ${\bf{p}}_\perp$ in the perpendicular x- and y-direction combined with Lorentz symmetry and express all eigenvalues through the set $\{\lambda\}$ at ${\bf{p}}_\perp=0$ (see Ref.~\cite{Nickel:2009wj} for details), giving
\begin{eqnarray}
\label{eq:Omega3}
\Omega(T,\mu;M(z))
&=&
-\frac{2T N_c}{V_\parallel}
\sum_{\lambda}
\int\frac{d{\bf{p}}_\perp}{(2\pi)^{2}}
\ln\left(2\cosh\left(\frac{\lambda\sqrt{1+{\bf{p}}_\perp^2/\lambda^2}-\mu}{2T}\right)\right)
+
\frac{1}{V}\int_V
\frac{( M({\bf{x}})-m)^2}{4G_s}
\nonumber\\&&
+
\mathrm{const.}
\,.
\end{eqnarray}
Second, we observe that the Hamiltonian ${H}_{MF}$ for the case ${\bf{p}}_\perp=0$ can be cast into the form
\begin{eqnarray}
H_{MF;1D}
&=&
\left(
\begin{array}{cc}
H_{1D} & \\
& H_{1D}
\end{array}
\right)
\,,
\end{eqnarray}
where $H_{MF;1D}$ is the Hamiltonian of the GN model for the same order parameter $M(z)$.
Since the spectral densities for the considered inhomogeneous phases are analytically known, we can evaluate the thermodynamic potential and in particular the sum $\sum_\lambda$ over  the eigenvalue spectrum.
Using the gap-equation it can furthermore be shown that self-consistent solutions, i.e. local minima of the thermodynamic potential, in the GN model can generically be used to find self-consistent solutions in the NJL model~\cite{Nickel:2009wj}.

\begin{figure}
\includegraphics[width=7.cm]{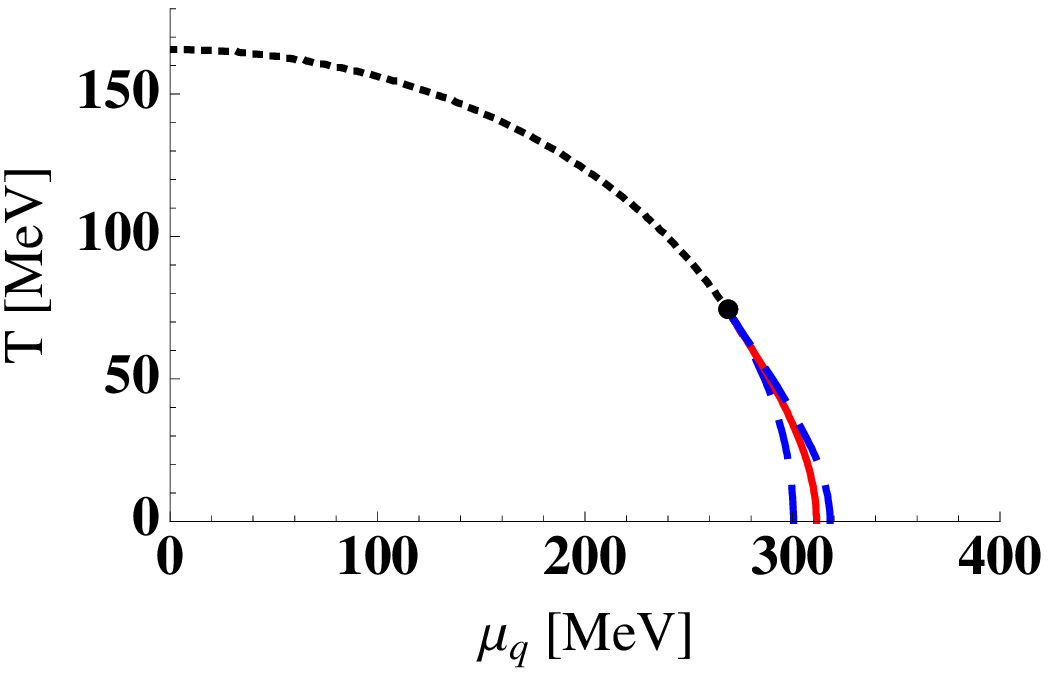}
\hspace{1cm}
\includegraphics[width=7.cm]{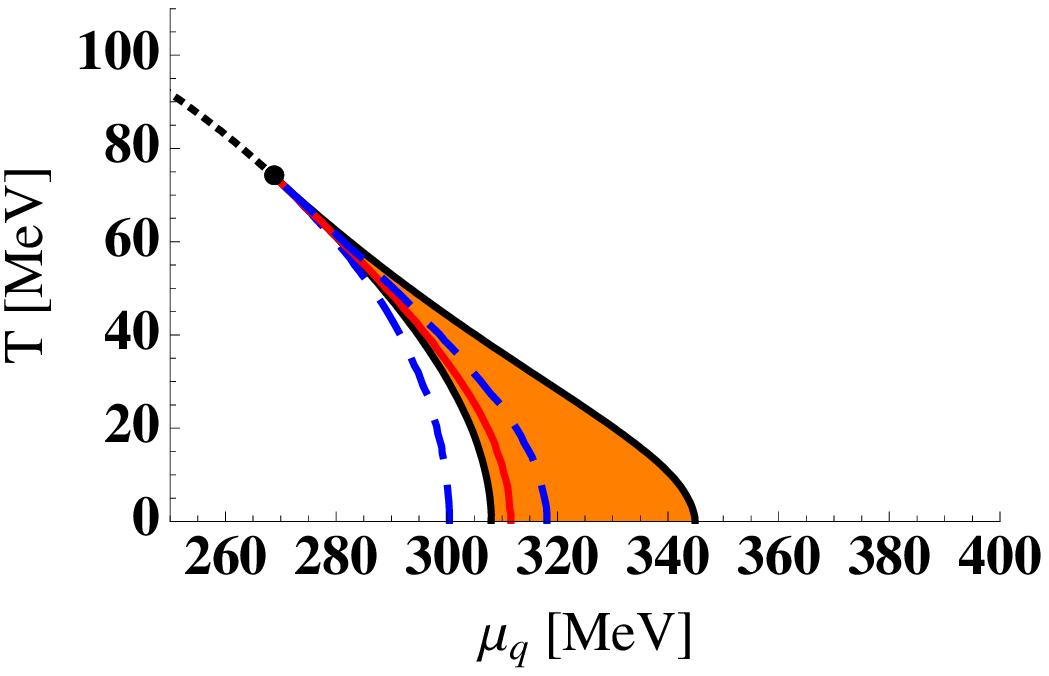}
\caption{
Left:
Structure of the NJL phase diagram in the chiral limit as a function of temperature $T$ and quark chemical potential $\mu_q$ for $M_q=300\mathrm{MeV}$.
The black (short-dashed) line indicates the second order phase transition from chirally broken to restored phase, the red (solid) line the first order phase transition and the bullet the critical point.
The spinodal region is enclosed by the blue (long-dashed) lines.
Right:
Same plot as on the left including the orange (shaded) domain where the energetically preferred ground state is inhomogeneous.
}
\label{fig:fig2}
\end{figure}

\begin{figure}
\includegraphics[width=7.cm]{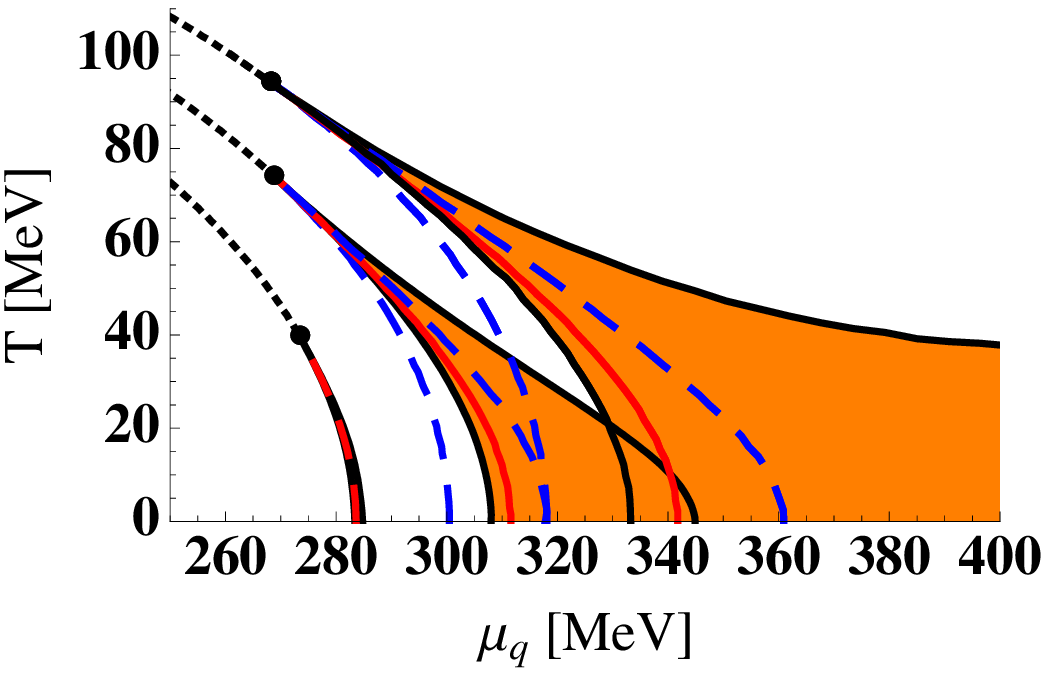}
\hspace{1cm}
\includegraphics[width=7.cm]{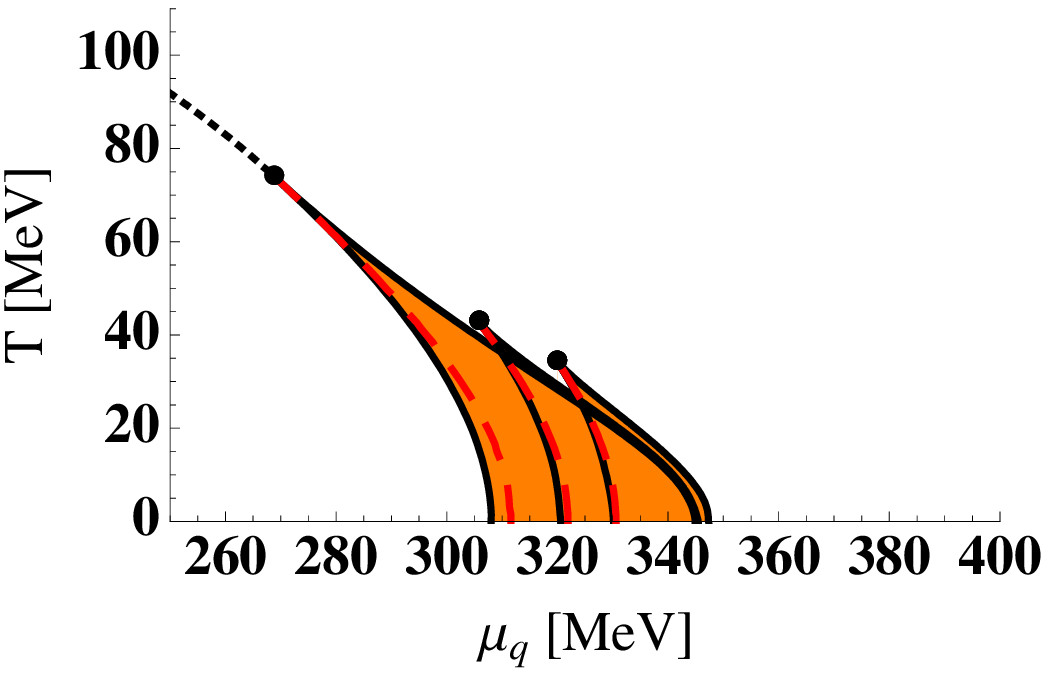}
\caption{
Left: Same plot as on the right of Fig. 2, now including results for $M_q=350\mathrm{MeV}$ (upper branch) and $M_q=250\mathrm{MeV}$ (lower branch).
Right: Same plot as on the right of Fig. 2, now including the domain of inhomogeneous phases for $m=5\mathrm{MeV}$ and $m=10\mathrm{MeV}$. Branches with critical points at smaller temperature $T$ and larger quark chemical potential $\mu_q$ correspond to larger current quark masses $m$.
}
\label{fig:fig3}
\end{figure}

\noindent
Consequently, it is possible to use the analytically known solutions in the 1+1 dimensional GN model for exploring the NJL  phase diagram including inhomogeneous phases with a one-dimensional modulation.
The form of the order parameter  of interest here has the specific form~\cite{Schnetz:2005ih}
\begin{eqnarray}
\label{eq:MzReal}
M(z)
&=&
\Delta
\left(
\nu\,
\mathrm{sn}(b\vert \nu)
\mathrm{sn}(\Delta z\vert \nu)
\mathrm{sn}(\Delta z+b\vert \nu)
+
\frac{
\mathrm{cn}(b\vert \nu)\mathrm{dn}(b\vert \nu)
}{
\mathrm{sn}(b\vert \nu)
}
\right)
\,,
\end{eqnarray}
where $\Delta$ is a scale parameter and $\mathrm{sn}$, $\mathrm{cn}$, $\mathrm{dn}$ are elliptic Jacobi functions with elliptic modulus $\sqrt{\nu}$. 
Physically, it describes lattices of equidistant domain-wall solitons and it is worth noting that it also parametrizes self-consistent solutions at finite current quark masses $m$.

\noindent
As already noted in the context of the GL expansion, we have to introduce a regularization scheme in order to make a specific calculation of a phase diagram. Since the usually employed regularization schemes in momentum space rely on a conserved three-momentum for the quasi-particles and therefore on homogeneous phases, we apply a proper-time regularization and refer to Ref.~\cite{Nickel:2009wj} for details.
We fix the coupling constant $G_s$ and cutoff scale $\Lambda$ by choosing the pion-decay constant $f_\pi=88$MeV for $m=0$ and the constituent quark mass in the vacuum $M_q=250,300,350$MeV.
Minimizing the thermodynamic potential in the parameters $\Delta$, $\nu$ and $b$ for each $\mu$ and $T$, we obtain the phase diagrams shown in Fig.~\ref{fig:fig2} and~\ref{fig:fig3}.

\noindent
As suggested by the generalized GL expansion we find that the first order phase transition is replaced by two second order phase transitions enclosing a domain of inhomogeneous phases.
The location of the CP depends on the model parameters and the region of inhomogeneous phases seems strongly correlated with the strength of the first order phase transition when limiting to homogeneous phases.
To illustrate the latter we also show the extend of the spinodal region, where the thermodynamic potential possesses more than one local minimum.
The picture stays qualitatively the same when including finite current quark masses, only the second order phase transition in the case of homogeneous phase transition is rendered a cross-over here.

\section{Inhomogeneous phases in the QM model}
\noindent
The NJL model regularized by a proper-time regularization and adjusted to chiral condensate and pion decay constant is known to give constituent quark masses of order $200\mathrm{MeV}$ in the vacuum. Hence it gives an undesired phenomenology with regard to the QCD phase diagram, mainly because quasi-particles will start forming a Fermi surface at $\mu\simeq M_q$. This is phenomenologically unacceptable for $\mu< (M_N-B)/3\simeq308\mathrm{MeV}$, where $M_N$ is the nucleon mass and $B$ the binding energy of nucleons in nuclear matter.
In the study of the phase diagram we have therefore chosen to fix $f_\pi$ to its phenomenological value and varying the value of $M_q$.
Since we haven't found a regularization that avoids these problems, we also discuss a model that is very similar to the NJL model and where the issue of the regularization scheme can be surpassed: The linear sigma model, which in this context is usually named QM model~\cite{Scavenius:2000qd,Schaefer:2006ds}.

\noindent
The Lagrangian of the QM model with $N_f=2$ and $N_c=3$ is given by
\begin{eqnarray}
\mathcal{L}_{QM}
&=&
\bar{\psi}
\left(
i\gamma^\mu \partial_\mu
-
g(\sigma+i\gamma_5 \tau^a\pi^{a})
\right)
\psi
-
U(\sigma,\pi^{a})
\,,
\nonumber\\
U(\sigma,\pi^{a})
&=&
-
\frac{1}{2}
\left(
\partial_\mu\sigma \partial^\mu\sigma 
+
\partial_\mu\pi^{a} \partial^\mu\pi^{a}
\right)
+
\frac{\lambda}{4}
\left(
\sigma^2
+
\pi^{a}\pi^{a}
-
v^2
\right)^{2}
-
c\sigma
\,,
\end{eqnarray}
$\psi$ is again the $4N_f N_c$-dimensional quark spinor, $\sigma$ the scalar field of the $\sigma$-meson and $\pi^{a}$ the pseudo-scalar fields of the pion triplet.
In mean-field approximation we treat the fields $\sigma$ and $\pi^{a}$ as classical and replace them by there expectation values~\cite{Scavenius:2000qd,Schaefer:2006ds}.
Furthermore we can use low-energy relations to connect the parameters $c$, $g$, $\lambda$ and $v^2$ with hadronic observables.
We will express those by the pion-decay constant $f_\pi$, the constituent quark mass in the vacuum $M_q$, the pion mass $m_\pi$ and $\sigma$-meson mass $m_\sigma$ via
$\langle\sigma\rangle=f_\pi$,
$\langle\pi^a\rangle=0$,
$c=m_\pi^2 f_\pi$,
$g=M_q/ f_\pi$,
$\lambda=(m_\sigma^2 - m_\pi^2)^2/ (2f_\pi^2)$ and
$v^2=f_\pi^2 - m_\pi^2/ \lambda$.

\noindent
For the thermodynamic potential in mean-field approximation we only include the contributions of the fermionic fluctuations and approximate
\begin{eqnarray}
\label{eq:Omega1QM}
\Omega_{QM}(T,\mu;\sigma({\bf x}))
&=&
-
\frac{T N_c}{V}
\sum_{n}
\mathrm{Tr}_{D,f,V} \, \mathrm{Log}\left(\frac{1}{T}\left(i\omega_{n}+\tilde{H}_{MF,QM}-\mu\right)\right)
+
\frac{1}{V}\int_V
U(\sigma({\bf x}),\pi^a({\bf x}))
\,,
\nonumber\\
\end{eqnarray}
where $\sigma({\bf x})$ is taken to be the only non-vanishing expectation value and the Hamiltonian reads
$
\tilde{H}_{MF,QM}
=
-i\gamma^0\gamma^{i}\partial_{i} + \gamma^0 g\sigma({\bf x})
$.
With the identification $M({\bf x})=g\sigma({\bf x})$ we can therefore evaluate the functional trace-logarithm for the same inhomogeneous phases as in the case of the NJL model.
In general those phases need not to be self-consistent solutions of the QM model, but for the purpose of checking whether an inhomogeneous phase is preferred compared to any homogeneous phase, this approach is sufficient.

\begin{figure}
\includegraphics[width=7.cm]{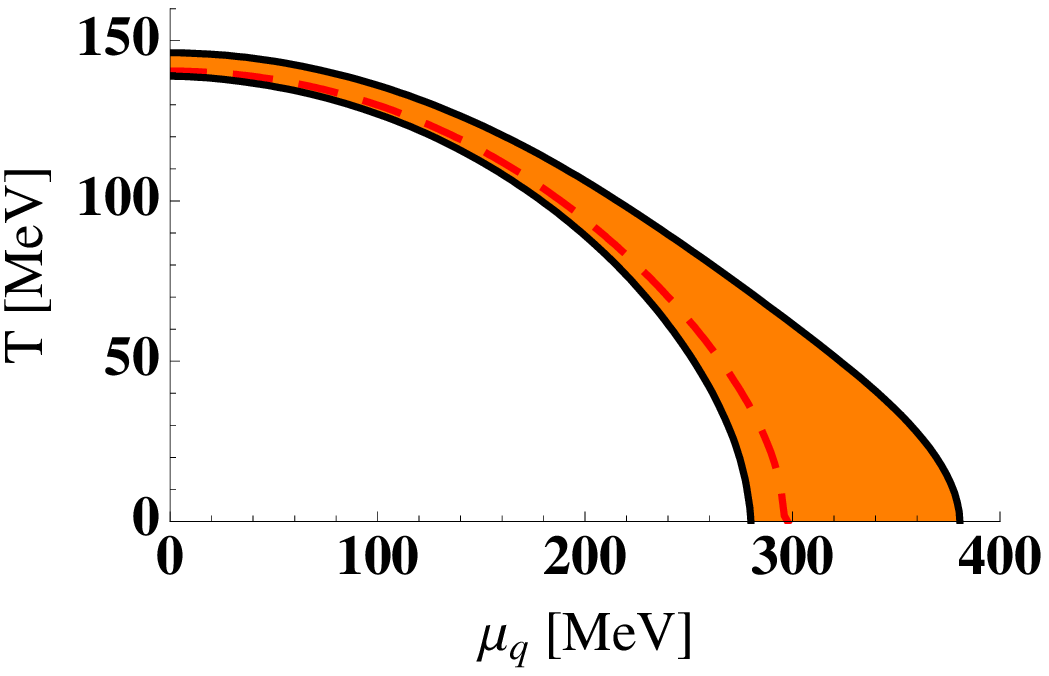}
\hspace{1cm}
\includegraphics[width=7.cm]{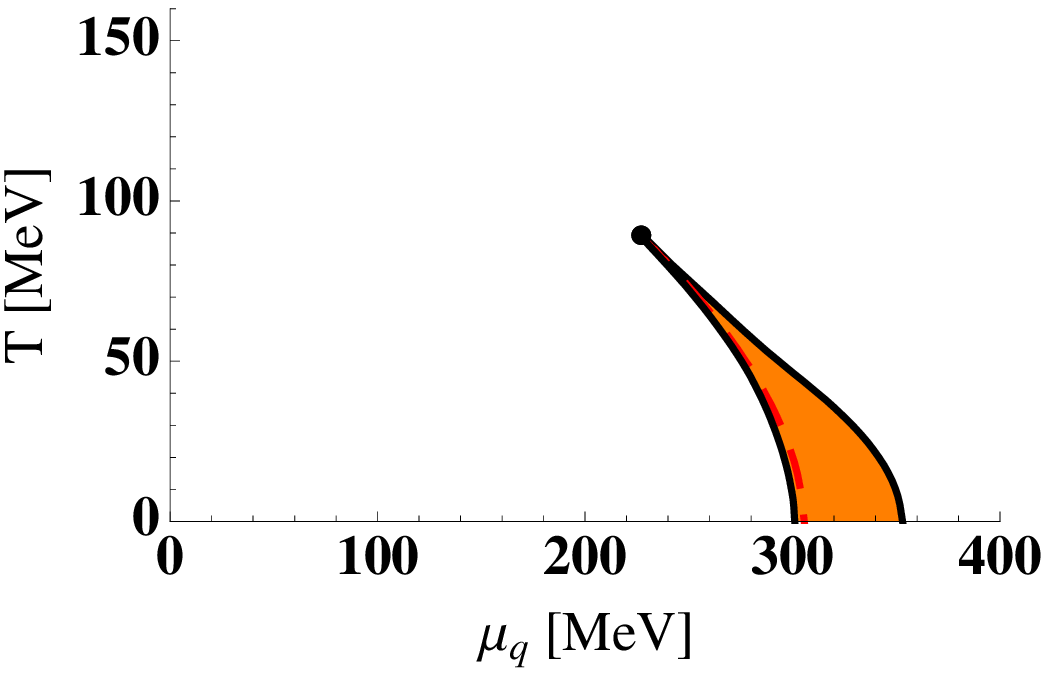}
\caption{
Left: Phase diagram for the QM model in the chiral limit with the red (dashed) line indicating the first order phase transition when limiting to homogeneous phases only. Inhomogeneous phases are preferred in the orange (shaded) domain enclosed by black (solid) lines.
Right: Same plot as on the left, now for $m_\pi=138\mathrm{MeV}$ and with the bullet showing the critical point.
}
\label{fig5}
\end{figure}

\noindent
The QM model is renormalizable which means that the divergences in the functional trace-logarithm can in principle be absorbed by the model parameters.
Instead of a proper renormalization we will however follow Refs.~\cite{Scavenius:2000qd,Schaefer:2006ds}, where it has been assumed that the zero temperature contribution can well be approximated by $\frac{1}{V}\int_V U(\sigma({\bf x}),\pi^a({\bf x}))$ with the parameters directly adopted to pheno\-menology.
Choosing the model parameters through
$f_\pi=93\mathrm{MeV}$, $M_q=300\mathrm{MeV}$, $m_\sigma=600\mathrm{MeV}$ and 
considering the chiral limit $m_\pi=0$ as well as the physical point with pion mass $m_\pi=138\mathrm{MeV}$, we obtain the phase diagrams shown in Fig.~\ref{fig5}.
The model in the applied approximation does not have a critical point in the chiral limit and, as maybe expected from the results obtained in the NJL model,  the first order phase transition from the chirally broken to the restored phase is replaced by two second order phase transitions  enclosing a domain where inhomogeneous phases are preferred.
For the physical value of the pion mass, we find a CP and the qualitative structure of the phase diagram is the same as in the NJL model.

\acknowledgments
\noindent
We thank G. Basar, M. Buballa, S. Carignano, G. Dunne, R. Pisarski, K. Rajagopal, B.-J. Schaefer and M. Stephanov for helpful comments and discussions.\\
This work was supported in part by funds provided by the U.S. Department of Energy (D.O.E.) under cooperative research agreement DE-FG0205ER41360 and by the German Research Foundation (DFG) under grant number Ni 1191/1-1.

\end{document}